\documentclass[12pt]{article}
\usepackage{amssymb,amsmath,amscd,epsf}
\usepackage[utf8]{inputenc}
\usepackage[english]{babel}

\usepackage[dvipsnames]{xcolor}
\usepackage[bookmarks=false]{hyperref}
\hypersetup{
    colorlinks=true,
    linkcolor=Sepia,
    citecolor=Mahogany,
    urlcolor=Blue, %cyan,
}

\usepackage{tikz}

\newcommand{\D}{\mathcal D}
\newcommand{\Z}{\mathcal Z}
\newcommand{\Tr}{\text{Tr}}
\newcommand{\tr}{{\rm tr}}
\newcommand{\Det}{{\rm Det}}
\newcommand{\Spec}{{\rm Spec}}

\newcommand{\ii}{{\mathrm i}}

\newcommand{\onehalf}{\mbox{{\large$\frac{1}{2}$}}}

\newcommand{\YM}{Yang-Mills}
\newcommand{\SDYM}{self-dual \YM\ theory}
\newcommand{\mn}{{\mu\nu}}
\newcommand{\lk}{{\lambda\varkappa}}

\newcommand{\cd}{\!\cdot\!}

\newcommand{\Mi}{M_i^{\,2}}
\newcommand{\NPV}{N_{PV}}
\newcommand{\PV}{Pauli-Villars}

\makeatletter
\let\@fnsymbol\@alph
\makeatother

\begin{document}

\title{Ultraviolet Properties of the Self-Dual Yang-Mills Theory}
\author{
Andrey Losev
\thanks{Federal Science Centre "Science Research Institute of System Analysis at Russian Science Academy" 
\hspace*{8mm}(GNU FNC NIISI RAN), Moscow, Russia}
\thanks{National Research University Higher School of Economics (HSE), Moscow, Russia}
\thanks{Moscow Institute of Physics and Technology (MIPT), Dolgoprudnyi, Russia}
\and
Igor Polyubin
\thanks{ Landau Institute for Theoretical Physics (ITP), Chernogolovka, Russia}
\thanks{Alikhanov Institute for Theoretical and Experimental Physics (ITEP), Moscow, Russia} 
\footnotemark[3] 
\and 
Alexei Rosly
\footnotemark[5] 
\footnotemark[2] 
\thanks{Institute for Information Transmission Problems (IITP), Moscow, Russia} 
\thanks{Skolkovo Institute of Science and Technology (Skoltech), Moscow, Russia}
\footnotemark[3] 
}
\date{February 07, 2018}

\maketitle

\begin{abstract}
    We compute the ultraviolet divergences in the \SDYM, both in the purely perturbative 
    (zero instanton charge) and topologically non-trivial sectors. It is shown in particular that 
    the instanton measure is precisely the same as the one-loop result in the standard \YM\ theory.
\end{abstract}

\section{\large Introduction} 
In this paper we shall study a four dimensional gauge theory sometimes called Chalmers-Siegel theory \cite{CS} 
or self-dual \YM\ theory, which is described by the following action functional:
\begin{equation}\label{S}
S=-\ii\!\int d^4x\;\tr\, P^{\mn} F_{\mn} - \frac{\tau}{16\pi^2}\int d^4x\;\tr\, F_{\mn}\tilde F^{\mn} \,,
\end{equation}
where $F_{\mn}$ is the field strength of a gauge field $A_\mu$\,, $F=dA+\frac{1}{2}[A\wedge A]$, for the gauge group $G=SU(N)$. 
The second term is a topological term with 
$\tilde F^{\mn}=\frac{1}{2}\varepsilon^{\mn\lk}F_\lk$\,, or, simply $\tilde F=*F$, 
and $\tau$ is a constant. 
A second fundamental field in this theory, $P_\mn$\,, is an $su(N)$-valued anti-self-dual antisymmetric tensor, 
$P=-*P$. The trace, tr, is understood in the $N$-dimensional representation. 
The same action can be conveniently written as
\begin{equation}
S=2\ii\!\int\tr\, P\wedge F - \frac{\tau}{8\pi^2}\int\tr\, F\wedge F \,.
\end{equation}
The latter form is applicable on an arbitrary four-dimensional Riemann manifold. 
We shall consider this field theory only in Euclidean signature. The field equations obviously are
\begin{equation}\label{fe1}
F=*F \,,
\end{equation}
\begin{equation}\label{fe2}
\nabla P=0 \,,
\end{equation}
where $\nabla =d+A$ is the exterior covariant derivative. We have thus the non-linear self-duality equation for a gauge field 
and a linear anti-self-dual field propagating in a background of a non-abelian instanton. 
One can think of field equations (\ref{fe1}),(\ref{fe2}) as describing a \YM\ field whose anti-self-dual ``part" is linearized, 
while the self-dual ``part" remains fully non-linear. These words can be given a more precise sense as was described 
in the literature, see \cite{CS,C1,C2}.

An attempt to transfer the above definitions to Minkowski space faces an immediate obstruction as the (anti) self-duality condition 
in a Lorentzian signature requires complex fields. As a result, there is apparently no way to give a physical meaning to the 
theory in view. Nevertheless, it might be interesting to note that, proceeding in perturbation theory formally with complex fields 
governed by the action functional (\ref{S}), one can compute scattering amplitudes with the following outcome. The only 
non-vanishing amplitudes will be the one-loop amplitude with an arbitrary number of external particles (``gluons") 
which have to be of the same helicity, say +1. These positive helicity\footnote{We do not go into details 
of conventions relating +/- helicity with anti/self-duality, which also depend on whether we speak of in- or out-particles.} 
``gluons" correspond to quanta of the non-linear gauge field $A_\mu$\,, 
while the quanta of $P_\mn$ should be viewed as having helicity -1.
Note, that all trees as well as all the opposite, minus helicity one-loop amplitude 
vanish, whence there is no place for $CPT$-inavriance and unitarity. The fact that the Lagrangian (\ref{S}) 
reproduces the ``all plus'' amplitudes was 
understood long ago, see \cite{CS,C1,C2}. Let us also mention that the amplitudes with particles of the same 
helicity always vanish in any supersymmteric theory. Thus, a supersymmetric version of (\ref{S}) 
would be a theory with absolutely no amplitudes.
As a matter of fact, even the off-shell one-loop diagrams vanish in the supersymmetric version (they correspond to $F$-terms), 
thus leaving us with a quantum theory equivalent to its classical limit, see, however, refs.\ \cite{FLN1,FLN2} where 
certain observables in supersymmetric SDYM were constructed obeying non-trivial conformal properties. 
Let us also mention the work in ref.\ \cite{K} in the direction of extending the study of the \SDYM\ 
to the case of Self-Dual Gravity. 

Since the on-shell content of our (non-supersymmteric) field theory is so poor we turn to studying some off-shell quantities. 
We shall see that 
despite its rudimentary character this theory shares certain basic features with other, more respectable QFT's. 
Moreover, it is tempting to conjecture that the \SDYM\ (SDYM for short) is in fact a conformal quantum field theory, 
providing us with a may be simplest non-trivial example of a CFT in four dimensions which has no supersymmetry. 
At present we do not however have sufficient arguments for this.

\section{\large Perturbative Self-Dual Yang-Mills Theory} \label{sect2}

The first crucial feature of the SDYM is that a would-be coupling constant 
in front of the first term in eq.\ (\ref{S}) is in fact an inessential parameter. Indeed, it can be made 
arbitrary by rescaling the field $P_\mn$\,. Therefore, the quantum SDYM theory has 
no {\em running} coupling constant. This happens despite the ultraviolet divergences, which are present in the SDYM 
and are very much similar in structure to the ordinary \YM\ theory as we shall see in the next section. 
Nevertheless, the absence of a running coupling constant should be taken as a sign of a (properly understood) 
conformal invariance of the theory. 

A first consequence of the absence of a coupling constant that can be drawn immediately is that the perturbation theory 
has very few terms. More precisely, as one can easily convince oneself, the only non-vanishing connected correlation 
functions in the SDYM are tree and one-loop ones, namely:
\begin{equation}\label{AP}
\langle A(x_1)P(x_2)\dots P(x_n)\rangle^{\rm connected}_{\rm tree} \,,
\end{equation}
\begin{equation}\label{PP}
\langle P(x_1)\dots P(x_n)\rangle^{\rm connected}_{\rm 1-loop} \,.
\end{equation}
An operator $A_\mu(x)$ in eq.\ (\ref{AP}) is understood in terms of a gauge-fixed Faddeev-Popov ghost formalism, 
which will explicitly appear later. Note that the above $n$-point off-shell-functions can be subject to reduction 
formulas (amputating legs and putting on-shell) to provide ``amplitudes" mentioned earlier. Upon this procedure, 
the 1-loop function of eq.\ (\ref{PP}) will give the ``all plus'' 1-loop amplitude, which we spoke about in the last section. 
Subject to the same procedure, the $n$-point function of eq.\ (\ref{AP}) gives zero, for it leads to a tree with 
one particle of helicity minus and all others of helicity plus. Such a tree amplitude is known to vanish 
in the standard \YM\ theory \cite{PT}. 
Here, exactly the same happens. 

We proceed to investigating the ultraviolet divergencies in the \SDYM\ theory. To start with, let us first notice 
what could be other local terms in the action functional which would have the same scaling dimension as those 
in the Lagrangian of eq.\ (\ref{S}). Obviously there are two main possibilities:
\begin{equation}
\tr P^\mn P_\mn {\rm ~~~and~~~} \tr F^\mn F_\mn \,.
\end{equation}Adding the first term would mean a complete change of the content of the theory: with $\tr P\cd P$ added 
to the action (\ref{S}) the theory becomes a full-fledged \YM\ theory, which is readily seen by performing 
a gaussian integration over $P_\mn$\,. On the other hand, adding a term $Z\tr F\cd F$ for some constant $Z$ does not 
essentially change anything in the \SDYM\ Lagrangian. Indeed, we can write 
\begin{equation}\label{FF=}
F\cd F\equiv F^\mn F_\mn = 2F^-\cd F^- + F\cd\tilde F \,, 
\end{equation}
where $F^-=\frac{1}{2}(F-\tilde F)$ is the anti-self-dual part of the field strength $F_\mn$\,. 
The term proportional to $F^-\cd F^-$ can be absorbed into the $P\cd F=P\cd F^-$ term in the action by 
redefining the anti-self-dual field $P$, while the second term in eq.\ (\ref{FF=}) yields a correction to 
the topological term in the action.

It can be immediately seen that the quantum corrections do not generate a (counter) term like $\tr P\cd P$, 
thus leaving the theory self-consistent. Indeed, the only one-particle irreducible diagrams have the field 
$A_\mu$ on the external legs. This follows from the above discussion around eq.\ (\ref{PP}) 
by noticing that upon amputation $P$ becomes $A$. Among these OPI diagrams the following ones 
contain ultraviolet divergences:
\begin{equation}\label{3graphs}
\begin{tikzpicture}
\draw (0,0) circle (5mm);
\draw (3,0) circle (5mm);
\draw (6,0) circle (5mm);
\fill (-0.5,0) circle (1.5pt);
\fill (0.5,0) circle (1.5pt);
\draw (-0.7,0) to (-0.5,0);
\draw (0.5,0) to (0.7,0);
\fill (3,-0.5) circle (1.5pt);
\fill (2.57,0.25) circle (1.5pt);
\fill (3.43,0.25) circle (1.5pt);
\draw (3,-0.5) to (3,-0.7);
\draw (2.57,0.25) to (2.32,0.45);
\draw (3.43,0.25) to (3.68,0.45);
\fill (5.5,0) circle (1.5pt);
\fill (6.5,0) circle (1.5pt);
\fill (6,0.5) circle (1.5pt);
\fill (6,-0.5) circle (1.5pt);
\draw (5.5,0) to (5.3,0);
\draw (6.5,0) to (6.7,0);
\draw (6,0.5) to (6,0.7);
\draw (6,-0.5) to (6,-0.7);
\end{tikzpicture}
\end{equation}
where the latter two are related to the first one by gauge invariance. 
These divergences should be compensated by $F^2$ type counterterms. 
In the topologically trivial case, when $\int\tr F\cd\tilde F=0$, we cannot distinguish between 
$\int\tr F\cd F$ and $\int\tr F^-\cd F^-$. However, in Sect.\ \ref{sect3}, we shall see that in 
general both terms appear in the bare action. Namely, the required counterterms can be written as 
\begin{equation}\label{cnt}
\begin{split}
    & \onehalf Z_-\,\tr F^-\cd\, F^- \; + \;\; \tilde Z\,\tr F\cd\tilde F \,, \\
    & Z_-=-\gamma_-\log M/\mu  \,, \;\;\;  \tilde Z=\beta\log M/\mu \,,
\end{split}
\end{equation}
where $M$ is a regulator mass, $\mu$ a renormalization scale, and $\gamma_-$ and $\beta$ are 
some constants whose numerical values will be computed below. 
The appearance of the first term has the effect that the field $P$ gets renormalized by a shift:
\begin{equation}\label{redefP}
P\;\to\;P+\frac{\ii}{2}\, Z_-\,F^- \,,
\end{equation}
while the topological term in eq.\ (\ref{S}) gets a correction from the second term in eq.\ (\ref{cnt}): 
\begin{equation}
    \tau\,\to\,\tau-16\pi^2\tilde Z \,.
\end{equation}
Note that the `topological coupling' $\tau$ starts running, but its imaginary part, the $\theta$-angle, 
remains unrenormalized. 
Note also that the renormalization (\ref{redefP}) with $Z_-$ being a real renormalization constant, pushes 
the field $P$ into the complex domain. In a sense, one can speak of a renormalization of the integration contour 
in the path integral. Moreover, since $P$ is shifted by $F^-$, which can be written as 
$P\to P - Z_-\frac{\delta S}{\delta P}$\,, 
it might be appropriate to recall the gradient flow approach to the study of oscillating integrals \cite{A}. 
This might deserve a further speculation, which we however postpone till better times.

To find the constants $\gamma_-$ and $\beta$ we have to perform the computation of the UV divergences 
in a topologically non-trivial background. This is done in the next section. However, even 
before that it might be interesting to consider the renormalization of Green's functions in the 
zero instanton case, that is a purely perturbative contribution to correlators (\ref{PP}). 
Note that the perturbation theory terminates in our case at one loop. Let us consider a two-point function, 
\begin{equation}
G_{\mn,\lk }(x,y):=\langle P_\mn(x)P_\lk(y)\rangle_o \,,
\end{equation}
where we omit the superscripts corresponding to adjoint representation; 
the subscript "$o$" indicates that this Green's function is computed with topologically 
trivial boundary conditions at infinity. The renormalization group equation determined by the 
expression (\ref{cnt}) and/or (\ref{redefP}) takes the following from: 
\begin{equation}
\mu\frac{\partial}{\partial\mu}\langle P(x)P(y)\rangle_o=
\frac{\ii}{2}\gamma_-\langle F^-(x)P(y)\rangle_o + \frac{\ii}{2}\gamma_-\langle P(x)F^-(y)\rangle_o \,, 
\end{equation}
where $\gamma_-=\partial Z_-/\partial\log\mu$\,; but 
\begin{equation}\label{cntct}
\langle F^{a\,-}_\mn(x)P^b_\lk(y)\rangle = 2\ii\,\delta^{ab}\,\delta (x-y)\,\Pi^-_{\mn,\lk} 
\end{equation}
is a pure contact term, because 
\begin{equation}
    F^{a\,-}_\mn=2\ii\:\frac{\delta S}{~\delta P_\mn^a}\;, 
\end{equation}
and we use a basis in the Lie algebra, such that $\tr P\cd F= \onehalf\sum_a P^{a\,\mn}F^a_\mn$\,.
The $\Pi^-$ in 
eq.\ (\ref{cntct}) denotes a projector to anti-self-dual anti-symmetric tensors: 
\begin{equation}\label{Pi-}
\Pi^-_{\mn,\lk} = \frac{1}{4} \big(g_{\mu\lambda}g_{\nu\varkappa}-g_{\mu\varkappa}g_{\nu\lambda}-
\varepsilon_{\mn\lk}\big) \,,
\end{equation}
which obeys
\begin{equation}
\begin{split}
    \Pi^-_{\mn,\lk} & = \Pi^-_{\lk,\mn}=-\Pi^-_{\nu\mu,\lk} \;, \\ 
    \Pi^-_{\mn,\lk} & = - 
    \onehalf\,\varepsilon_{\mn\,\cdot\;\cdot}^{\phantom{\mn}\sigma\rho} \,
    \Pi^-_{\sigma\rho,\lk} \;.
\end{split}
\end{equation}
Thus, our Green function should depend on renormalization scale $\mu$ and satisfy
\begin{equation}\label{dmuG}
\mu\frac{\partial}{\partial\mu}G(x,y\,|\,\mu)=- 2\gamma_-\,\delta (x-y) \,.
\end{equation}
Here, $G$ is the same as $G_{\mn,\lk }$ with indices stripped off: $G_{\mn,\lk }=\Pi^-_{\mn,\lk}\cdot G$.

For separated points, $x\neq y$, one may expect on dimensional grounds that, for some constant $c$, 
\begin{equation}\label{G(xy)}
G_{\mn,\lk }(x,y)= c \:
\frac{\Pi^-_{\mn,\lk}}{\lVert x-y\rVert^4}\;,\;\;\; x\neq y \,,
\end{equation}
This cannot however be a complete description of $G(x,y)$, in first place because the singularity $\lVert x-y\rVert^{-4}$ 
is not integrable\footnote{We thank G.~Korchemsky for emphasizing this to us.}. 
Indeed, this singularity should be regularized by a contact term bringing in a $\mu$-dependence in accord with 
eq.\ (\ref{dmuG}). A natural ansatz for our two-point function is thus 
\begin{equation}\label{Gregn}
G_{{\text n.a.}}(x,y)=c\int\frac{d^4u}{u^4}\,[\delta(u+x-y)-e^{-\mu^2u^2}\delta(x-y)] \,,
\end{equation}
where we write $u^2$ instead of $\lVert u\rVert^2=u^\nu u_\nu$\,.
Computing $\partial/\partial\log\mu$ on the r.h.s. of eq.\ (\ref{Gregn})
and comparing with the renormalization group equation (\ref{dmuG}) we find that $\gamma_- =-\pi^2c$.
The choice of the exponential function 
$f(u)=\exp(-\mu^2u^2)$ is of course arbitrary to a certain extent; one needs only that $f(0)=1$ and that 
$f(u)/u^4$ is integrable at infinity. 
Different choices of $f(u)$ would mean a usual regularization ambiguity corresponding to 
finite local terms. With this understanding, in order to simplify some formulas, 
we may chose a contact term which looks slightly less natural, namely, 
\begin{equation}\label{Greg}
G(x,y)=-\frac{\gamma_-}{\pi^2}\int\frac{d^4u}{u^4}\,[\delta(u+x-y)-e^{\ii m\cdot u}\delta(x-y)] \,,
\end{equation} 
where $m$ is a 4-vector of norm squared $\lVert m \rVert^2=\mu^2$\,. We notice that there is in fact no 
breaking of $SO(4)$ invariance, for the integration projects to the invariant part of the integrand and 
the result depends only on $\mu$ and not on the vector $m$ itself. 
Note that, unlike $(x-y)^{-4}$\,, the expression in eq.\ (\ref{Gregn}) or (\ref{Greg}) is 
a well-defined 2-point distribution which possesses a well-defined Fourier image. 
In the latter case of eq.\ (\ref{Greg}), the Fourier image, 
$\tilde G(p,q)=\tilde G(p)\,\delta(p+q)$\,, 
is found in the following form: 
\begin{equation}\label{G(q)}
\tilde G(p) = \gamma_- \log(p^2/\mu^2) \,.
\end{equation}
In a traditional momentum-space computation of Feynman diagrams, 
the dependence on $\mu$ in eq.\ (\ref{G(q)}) comes from dealing with UV divergences of loop integrals. 
Here we saw that the same result can be obtained in real space by adjusting a contact term. 
This is essentially a simple particular case of the technique developed in ref.\ \cite{FJL}. 

Let us now confirm this result and find the value of $\gamma_-$ 
by directly computing the corresponding Feynman graphs in a standard way. 
We have to compute a one-loop (there are no higher loops) 2-point function corresponding to the first graph 
in (\ref{3graphs}). For this aim we write the gauge field as a sum $A+a$ with $A$ a classical background field 
and consider the quadratic with respect to the quantum field $a$ part of the Lagrangian. This is given, 
together with a choice of gauge-fixing and ghost terms, by the following expression: 
\begin{equation}\label{ql}
S=2\ii\!\int\tr(p\wedge\nabla a + b\,\nabla^*a) - \!\int\tr\,\bar c\,\nabla^*\nabla c \,,
\end{equation}
where $\nabla\! = d+A$ is the exterior covariant derivative in the classical background field $A_\mu$\,; 
the fields $a_\mu\,,\,p_\mn\,(p=-*p)$, scalar boson $b$, and scalar anti-commuting Faddeev-Popov 
ghosts $\bar c,c$ are quantum fields 
(all with values in $su(N)$); $\nabla^*=-*\nabla *$\,. 

In addition, we introduce \PV\ (PV) regulator fields with the following action functional:
\begin{equation}\label{PVL}
\begin{split}
S_{PV}=\ii\!\int\tr\big[\bar p_1\wedge\nabla a_1 + \bar b_1\,\nabla^*a_1 + 
\bar a_1\wedge\nabla p_1 + \bar a_1\,\nabla b_1\big] + \\ 
\int\tr\big[ N_{a_1}\,\bar a_1\cd a_1 + N_{p_1}(\bar p_1\cd p_1 + \bar b_1\cd b_1)\big] + \\
\ii\!\int\tr\big[p_2\wedge\nabla a_2 + b_2\,\nabla^*a_2\big] + 
\int\tr\big[ N_{a_2}\,a_2\cd a_2 + N_{p_2}(p_2\cd p_2 + b_2\cd b_2)\big] \,,
\end{split}
\end{equation}
where $a_1\,,\,p_1\,,\,b_1$ are complex anti-commuting partners to $a,\,p,\,b$, while 
$a_2\,,\,p_2\,,\,b_2$ are real commuting ones. 
The PV mass terms above need an explanation. The coefficients $N_{a_i}$ in front of $a_i^2$-terms have 
a natural scaling dimension of $M^2$, while the coefficients $N_{p_i}$ are dimensionless. 
As a matter of fact, only the products, 
\begin{equation}\label{MNN}
M_1^{\,2} = N_{a_1}\,N_{p_1} \,, \;\;\; M_2^{\,2} = N_{a_2}\,N_{p_2} \,,
\end{equation}
will appear in most formulas. Therefore, we can send $N_{a_i}$ to infinity and keep dimensionless parameters 
$N_{p_i}$ finite; they will not enter the resulting physical quantities. The choice of PV fields together with 
a quadratic relation on the PV masses $M_i$ must gaurantee a complete cancellation of the UV divergences. 
Introducing for convenience $M_0=0$ (a mass of `physical' fields), these relations can be written as follows: 
\begin{equation}\label{PVconstr}
\begin{split}
    (M_0)^0 - 2(M_1)^0 + (M_2)^0 = 0 \,, \\ 
    (M_0)^2 - 2(M_1)^2 + (M_2)^2 = 0 \,. 
\end{split}
\end{equation}
Furthermore, we have to introduce also \PV\ partners to the ghost fields, which we however do not 
write explicitly as they obey the same pattern as above. 

With these tools ready, we can compute the amputated 2-point function, a `polarization operator', 
cf., first graph in (\ref{3graphs}). For the gauge group $G=SU(N)$, the PV-regularized 
polarization operator can be written as
\begin{equation}\label{PsimN}
\Pi_\mn^{ab} = N\delta^{ab}\Pi_\mn(q\,|\,M) \,, 
\end{equation}
where the labels $a,b$ refer to the Lie algebra ${\mathfrak g}=su(N)$. 
Let us split $\Pi_\mn$ into two parts, 
\begin{equation}
\Pi_\mn = \Pi_\mn^{(apb)} + \Pi_\mn^{(gh)} \,, 
\end{equation}
corresponding to the contributions from the PV-supermultiplets of fields $a,p,b$ and $\bar c,c$\,, 
respectively. Then, the first term, for example, is given by the following momentum space integral: 
\begin{equation}\label{Papb}
\Pi_\mn^{(apb)}= 4\int \frac{d^4k}{(2\pi)^4} \sum_{i=0}^{\,2}f_i \, 
\frac{(k+q)_\mu k_\nu+k_\mu (k+q)_\nu -g_\mn\big[k\cd(k+q)+M_i^{\,2}\big ]}
{(k^2+M_i^{\,2})((k+q)^2+M_i^{\,2})} \,,
\end{equation}
where the sum runs over the PV degrees of freedom with 
\begin{equation}\label{fi}
f_0=1,\; f_1=-2,\; f_2=1 \,.
\end{equation}
There exist standard techniques for computing the integral in eq.\ (\ref{Papb}).
Fortunately, we do not have to really make a computation 
for the answer can be found in text-books. 
As a matter of fact, this one loop Feynman diagram coincides with a Dirac spinor loop in the QED. 
This coincidence could be foreseen {\it a priori} by observing that the fields $a,p,b$
in the Lagrangians (\ref{ql}),(\ref{PVL}) can be identified with the Dirac spinor fields via 
a `topological' twist \cite{W}. Indeed, let us consider two Dirac spinors $\Psi_j$\,, which, 
in Weyl notation, are written as $\psi_{\alpha j}\,,\psi_{\dot\alpha j}$\,, 
where $\alpha,\dot\alpha$ are Weyl spinor indices, and $j=1,2$. For the purposes of the present 
argument, the topological twist amounts to identifying the $R$-symmetry label $j$ with a 
dotted Weyl spinor index, say $\dot\beta$\,. Then, $\psi_{\alpha j}$ becomes a 4-vector 
$(a_\mu)=(a_{\alpha\dot\beta})$, while $\psi_{\dot\alpha j}\leadsto\psi_{\dot\alpha\dot\beta}$\,, 
split into its symmetric, 
$p_{\dot\alpha\dot\beta}:=\frac{1}{2}(\psi_{\dot\alpha\dot\beta}+\psi_{\dot\beta\dot\alpha})$, 
and skew-symmetric, $b:=\frac{1}{2}\varepsilon^{\dot\alpha\dot\beta}\psi_{\dot\alpha\dot\beta}$\,, 
parts, gets identified with an anti-self-dual tensor $p_\mn\,,\,p=-*p$, and a scalar $b$. 
Bearing this identification in mind, one can convince oneself that the polarization operator 
of eq.\ (\ref{Papb}) is precisely the same as in QED with (massive) Dirac spinor loop. 
Thus, we can read off the answer for the integral (\ref{Papb}) from text-books 
(e.g., \cite[Sect.\ 7.1.1]{IZ}). 
The contribution $\Pi_\mn^{(gh)}$ from the scalar (ghost) loops is a standard text-book 
result as well. All together, we obtain the following quantity: 
\begin{equation}\label{po}
\Pi_\mn(q\,|\,M) = - \frac{5}{3(4\pi)^2}\,
(q^2g_{\mu\nu}-q_{\mu}q_{\nu}) \log q^2/M^2 \,,
\end{equation}
where we set $M_1^{\,2}=2M^2,\, M_2^{\,2}=4M^2$, for convenience and 
to satisfy the constraint on the PV masses in eq.\ (\ref{PVconstr}). 

Going back to the unamputated Green function $\langle PP\rangle_o$\,, we have to apply 
inverse propagators to the polarization operator in eq.\ (\ref{po}). Then, we trade off 
the UV divergences, which manifest themselves in eq.\ (\ref{po}) with $M\to\infty$, for 
an emergent renormalization scale $\mu$, getting finally 
\begin{equation}
\langle P_\mn(p)P_\lk(q)\rangle_o = \Pi^-_{\mn,\lk}\, \tilde G(p)\,\delta(p+q) 
\end{equation}
with 
\begin{equation}\label{Gpol} 
\tilde G(p) = \frac{5}{3}\, \frac{N}{(2\pi)^2} \log p^2/\mu^2 \,.
\end{equation}
This coincides with eq.\ (\ref{G(q)}) 
and gives us the value of `anomalous dimension', \mbox{$\gamma_- = \frac{5}{3}\, \frac{N}{(2\pi)^2}$\,.} 
We have thus computed the first of two counterterms (\ref{cnt}). We shall confirm this 
as well as compute the second, topological, counterterm with help of heat kernel method 
in the next section.

It might be also worth noticing that the correlation function of the form (\ref{Greg}), (\ref{G(q)}) 
is also encountered in the standard \YM\ quantum theory. Namely, 
from the standard result on the gluon self-energy computed in the background field formalism, 
one can find the correlation function 
of anti-self-dual parts of the field strength (the Lie algebra labels are omitted):  
\begin{equation}
    \langle F_\mn^-(p)F_\lk^-(q)\rangle^{\text{YM\,1-loop}}_o = 
    - \Pi^-_{\mn,\lk}\, \bigg(1 - \frac{11}{3} \frac{g^2N}{16\pi^2} \log\frac{p^2}{\mu^2}
    \bigg)\,\delta(p+q) \,.
\end{equation}

\section{\large The Instanton Measure} \label{sect3}

A partition function of SDYM is a sum over instantons, that is a sum over topological sectors 
with non-negative instanton numbers. Let us consider a contribution from the fields of 
topological charge 
\begin{equation}\label{topch}
    \frac{1}{2(2\pi\ii)^2}\int\tr F\wedge F = k \geqslant 0 \,,
\end{equation}
where the gauge group is $G=SU(N)$, the gauge field and its field strength 
$F_\mn$ are anti-hermitean, and the trace is taken in the $N$-dimensional representation. This 
contribution is given by a measure on the $k$-instanton moduli space, which results from a 
perturbation theory in the background of a classical self-dual gauge field $A_\mu$\,. 
In the case at hand, the perturbation theory terminates at one loop. Therefore, everything 
can be described by the quadratic approximation defined in eq.\ (\ref{ql}) plus the \PV\ 
Lagrangian, see eq.\ (\ref{PVL}) and subsequent discussion. Thus, for the resulting PV-regulated 
measure we get 
\begin{equation}\label{IM1}
    dV_k = dv_k\, e^{-S_0} \frac{1}{\Z_0}
    \prod_{i=0}^{\NPV} D_i^{-f_i/2}\, C_i^{\,f_i} \,,
\end{equation}
where $S_0$ is the bare classical action, $dv_k$ the classical measure on the moduli space, 
which is exactly the same as in standard YM theory. 
$\Z_0$ is a normalization factor. 
The integer numbers $f_i$ in eq.\ (\ref{IM1}) 
play the same role as $f_i$'s defined in eq.\ (\ref{fi}) with some difference to be mentioned below. 
$D_i$ and $C_i$ are elliptic determinants, which will be specified shortly. The above measure can be 
used to compute the instanton contribution to a correlation function, for example, that of two fields 
$P$. Alternatively, one can exploit the background field method in the sector of $k>0$, which means that 
the background gauge field should not be precisely a self-dual instanton, but we should allow it to vary 
in a neighborhood of the latter. Then, for example, the contribution to amputated version of $\langle PP\rangle$ 
will be recovered from the second variation with respect to the background field as usual. We do not go 
into details of the formalism, because we are interested only in the structure of the UV-divergences. Thus, 
we consider the expression (\ref{IM1}), where the determinant factors, $D_i$ and $C_i$\,, contain 
the background field which is slightly different from an instanton. 

The $C_i$'s come from the ghost term 
in eq.\ (\ref{ql}) and corresponding PV-partners. Namely, 
\begin{equation}
    C_i = \Det (\Delta_{gh}+\Mi) \,,
\end{equation}
where $\Delta_{gh}$ is a scalar Laplacian, $\Delta_{gh}=\nabla^*\nabla$, in adjoint representation 
of the gauge group as in eq.\ (\ref{ql}), while the PV-masses are subject to certain constraints 
which ensure that the product $\prod C_i^{\,f_i}$ is finite for finite $\Mi$\,. Note that $\Delta_{gh}$ 
has no zero-modes in the background of an irreducible gauge field, which, in particular, requires 
$k\geq N/2$ (cf.\ \cite{AHS}), which we assume from now on. 

The other determinants in eq.\ (\ref{IM1}) are as follows: 
\begin{equation}
    D_0 = \Det'\, \D \,,
\end{equation}
where the prime means zero-modes removed, and $\D$ is a first order operator defined by the 
quadratic form for the fields $a,p,b$ in eq.\ (\ref{ql}). To describe it conveniently let us 
combine the field $p_\mn\,,\; p=-*p$, and scalar $b$ into a unified field $\hat p_\mn=p_\mn+g_\mn b$\,. 
Then, the quadratic form in eq.\ (\ref{ql}) is written as $\langle\Phi,\D\Phi\rangle$\,, where 
\begin{equation}\label{D}
    \Phi =  \begin{pmatrix} a \\ \hat p \end{pmatrix} \;, ~~~~~~~~
    \D = 
    \begin{pmatrix}
    0     & \ii D^* \\
    \ii D &   0 
    \end{pmatrix} 
\end{equation}
for $D,D^*$ to be explicitly given later, and 
$\langle\Phi,\Phi\rangle=\int\!\sqrt{g}\:\tr(a\cd a+\hat p\cd\hat p)$, where $g_\mn$ is a world-sheet metric 
and, e.g., $a\cd a=g^\mn a_\mu a_\nu$\,. The PV-partners are then
\begin{equation}\label{Di}
    D_j=\Det\,\D_j \;, ~~~~~~
    \D_j =
    \begin{pmatrix}
    N_{a_j}   & \ii D^* \\
    \ii D &   N_{p_j} 
    \end{pmatrix} 
\end{equation}
in accord with the PV-Lagrangian of eq.\ (\ref{PVL}). The real parameters $N_{a_j}\,,\,N_{p_j}$ 
were discussed in the previous section. Here again, the PV-masses $\Mi:=N_{a_i}N_{p_i}$ have to satisfy 
certain constraints ensuring that the product $\prod D_i^{-f_i/2}$ is finite for finite $\Mi$\,.

By saying that the product in eq.\ (\ref{IM1}) is finite for finite $\Mi$ we mean the following. Suppose 
we deal with a non-negative second order elliptic operator $A$ on a compact manifold. Then, we can define 
a product of determinants $\Det'(A+\Mi)$ by the following formula:
\begin{equation}\label{def}
  \sum_{i=0}^{\NPV} \log \Det'(A+\Mi) = 
    \int_0^\infty \frac{dt}{t}\; \sum_{i=0}^{\NPV} f_i\,\big(\Tr\, e^{-t(A+\Mi)}-n_0(A)e^{-t\Mi}\big) \,  ,
\end{equation}
where $M_0^{\,2}=0$, $\Det'A$ denotes a determinant with zero modes of the operator $A$ removed, 
and $n_0(A)$ denotes the number of those. $\Det'(A+\Mi)$ means similarly removing first $n_0(A)$ 
eigenvalues, which equal $\Mi$ each. The integral in eq.\ (\ref{def}) is convergent provided the integer 
parameters $f_i$ (meaning the number of degrees of freedom) and regulator masses $\Mi$ obey
\begin{equation}\label{0th}
\sum_{i=0}^{\NPV}f_i=0\,, ~~~ f_0=1 \,,\\
\end{equation}
\begin{equation}\label{2nd}
\sum_{i=1}^{\NPV} f_i\,\Mi=0 \,,
\end{equation}
\begin{equation}\label{4th}
\sum_{i=1}^{\NPV} f_i\,M_i^{\,4}=0 \,.
\end{equation}
Then, the ultraviolet divergence in $\Det'A$ manifests itself via the behavior of the quantity 
in eq.\ (\ref{def}) upon $\Mi\to\infty$. This behavior is determined by the short-time expansion 
of the heat kernel \cite{See}, $t\to+0$:
\begin{equation}\label{HK}
    \Tr\, e^{-tA}\, \sim\, \frac{\phi_2}{t^2}+\frac{\phi_1(A)}{t}+\phi_0(A)+O(t) \,.
\end{equation}
Here $\phi_2=\text{Vol}(X)/(4\pi)^2$, where $\text{Vol}(X)=\int\!\sqrt{g}$ is the volume of the manifold 
where the field theory is defined. This term of the expansion does not depend on the operator $A$. 
That is why we have not included a fourth order constraint on $M_i$\,'s in eq.\ (\ref{PVconstr}); 
the discussion 
in Sect.\ \ref{sect2} deals only with field-dependent 1-loop quantities. To satisfy more constraints 
on $\Mi$\,'s as in eqs.\ (\ref{0th}-\ref{4th}) one has to introduce more regulator fields. From now on 
we assume $\NPV$ to be large enough. 

One can then show that, for $\Mi\to\infty$, 
\begin{equation}\label{logdet}
    \sum_{i=0}^{\NPV} f_i\log \Det'(A+\Mi)\sim (\phi_0(A)-n_0(A)) \sum_{i=1}^{\NPV}f_i\log\Mi 
\end{equation}
provided certain further constraints are imposed to cancel divergences of types $M^2\log M^2$ and 
$M^4\log M^2$. Namely, we require that 
\begin{equation}\label{L2nd}
    \sum_{i=1}^{\NPV}f_i\,\Mi\log\Mi =0 \,,
\end{equation}
\begin{equation}\label{L4th}
    \sum_{i=1}^{\NPV}f_i\, M_i^{\,4}\log\Mi =0 \,.
\end{equation}
Finally, since the system of equations (\ref{2nd}),(\ref{4th}),(\ref{L2nd}),(\ref{L4th}) is invariant 
under a simultaneous rescaling of the masses we can set $\Mi=M^2\nu_i$ for some irrelevant dimensionless 
parameters $\nu_i$ and track only the large $M^2$ dependence. Then, we have for $M^2\to\infty$ that 
\begin{equation}
    \Det'A\,\prod_{i=1}^{\NPV}\Det(A+\Mi)^{f_i}\sim M^{-2\phi_0(A)}\, \tilde D \,.
\end{equation}
Note that the massive operators $A+\Mi$ got back their first $n_0(A)$ eigenvalues, which cancelled the 
term $-n_0(A)\sum f_i\log\Mi$ in the r.h.s.\ of eq.\ (\ref{logdet}). The last factor, 
$\tilde D$ does not depend on $M^2$. Its scaling dimension is $[{\rm mass}]^{2(\phi_0(A)-n_0(A))}$. 
This ensures that the exponential of the quantity in the l.h.s. of eq.\ (\ref{logdet}) 
is dimensionless, as it should be for $\sum_{i=0}^{\NPV}f_i=0$. 

With this understanding, let us now find out what are the coefficients $\phi_0$\,, which determine 
the large $M^2$ behaviour of the determinants in eq.\ (\ref{IM1}).
The ghost field contribution in eq.\ (\ref{IM1}) gives 
\begin{equation}\label{Dgh}
    \prod_{i=0}^{\NPV}C_i^{\,f_i} = \prod_{i=0}^{\NPV} \Det(\Delta_{gh}+\Mi)^{f_i} \sim 
    M^{-2\phi_0(\Delta_0)}\, \tilde C \,,
\end{equation}
where $\tilde C$ is a finite function of the instanton parameters, while the Seeley 
coefficient $\phi_0(\Delta_0)$ for the scalar Laplacian in adjoint representation can be 
found in \cite{Sch,RS,Sv,D}  
\begin{equation}\label{phigh}
    \phi_0(\Delta_0) = \frac{1}{12}\, \frac{2N}{(4\pi)^2} \int_X\sqrt{g}\,\tr\, F\cd F + \ldots \,,
\end{equation}
where the dots denote a contribution quadratic in the world-sheet curvature (cf.\ \cite{RS,D}). 
In the case of $X=S^4$, a four-sphere, this contribution is a constant, which does not depend on the 
radius of the sphere. 

The operator $\D$ in eq.\ (\ref{D}) governing the gauge field sector does have zero modes in the 
background we are dealing with. More precisely, for an irreducible background field of 
positive topological charge (\ref{topch}) we have that $n_0(D^*)=0$, while 
$n_0(D)=\text{index}(D)=4Nk-(N^2-1)$, which equals the number of instanton moduli. 
This is known from refs.\ \cite{AHS,Sch} and from the explicit form of the operators $D,D^*$ 
read off from eq.\ (\ref{ql}):
\begin{equation}\label{D,D*}
\begin{split}
    Da\, &= \Pi^-(\nabla a) + \nabla^* a \,, \\
    D^*\hat p &= \nabla^*\Pi^-  p  + \nabla b \,.
    \end{split}
\end{equation}
Here $\Pi^-=\onehalf (1-*)$ is the same projector to the anti-self-dual 2-forms as  in eq.\ (\ref{Pi-}),
and $Da$ is a sum of a 2-from $\Pi^-(\nabla a)$ and a 0-form $\nabla^* a$\,.

We shall consider below the corresponding Laplacian operators, $D^*D$ and $DD^*$, where, for instance, 
\begin{equation}\label{Lap}
D^*Da  =(\nabla^*\Pi^-\nabla +\nabla\nabla^*)a \,. 
\end{equation}
These two Laplacians are non-negative elliptic operators whose non-zero eigenvalue spectra coincide. 
Let us consider the eigensubspaces of $D^*D$ and $DD^*$ corresponding to a common eigenvalue $\lambda^2>0$. 
In an orthonormal basis, the operator in eq.\ (\ref{Di}) becomes then a matrix
\begin{equation}
    (\D_j)_\lambda = 
        \begin{pmatrix}
    N_{a_j}   & \ii\lambda \\
    \ii\lambda &   N_{p_j} 
    \end{pmatrix} \,, ~~~~~~~~
    \det (\D_j)_\lambda = \lambda^2 +\Mi \,.
\end{equation}
Therefore, we can formally write the elliptic determinant as a product:
\begin{equation}
    \Det'\,\D_i= \prod_{0\neq\lambda^2\in\Spec (D^*D)}(\lambda^2 +\Mi)= \Det'(D^*D+\Mi) \,.
\end{equation}
In the zero mode subspace of $D$, the operator $\D_i$ acts simply by a multiplication with 
$N_{a_i}=\Mi N_{p_i}^{-1}$. Hence, for $i=1,\ldots,\NPV$,
\begin{equation}\label{detDi}
    D_i\equiv \Det\D_i= \Det(D^*D+\Mi)\cdot (N_{p_i})^{-n_0(D)} \,,
\end{equation}
where the last factor is dimensionless and does not depend on $M^2$ when we set $M^2\to\infty$\,. 
The large $M^2$ behaviour of the determinants in eq.\ (\ref{detDi}) is described, according to 
the discussion above, by a Seeley coefficient $\phi_0(D^*D)$ and, for $M^2\to\infty$\,, we have for 
the gauge field sector of eq.\ (\ref{IM1}): 
\begin{equation}\label{Dgauge}
    \prod_{i=0}^{\NPV} D_i^{-f_i/2} = (\Det'\,D^*D)^{-1/2} 
    \prod_{i=1}^{\NPV} \Det(D^*D+\Mi)^{-f_i/2}\, \tilde N \sim 
    M^{\phi_0(D^*D)} \, \tilde D \,,
\end{equation}
where $\tilde N, \tilde D$ do not depend on $M^2$. 
The value of $\phi_0(D^*D)$ can be found from refs.\ \cite{RS,D}: 
\begin{equation}\label{phigauge}
    \phi_0(D^*D) = -\frac{2}{3}\, \frac{2N}{(4\pi)^2} \int_X\sqrt{g}\,\tr\, F\cd F 
    - 2 \frac{2N}{(4\pi)^2} \int_X \tr\, F\wedge F 
    + \ldots \,,
\end{equation}
where the dots denote a contribution, which becomes a constant in the case $X=S^4$. 
From the formulae (\ref{Dgh}) and (\ref{Dgauge}) we see that the one-loop divergence is 
given by $(\phi_0(D^*D)-2\phi_0(\Delta_0))\log M$, which should be cancelled by counterterms 
in the bare action:
\begin{equation}\label{59}
    S_0 = S + (\phi_0(D^*D)-2\phi_0(\Delta_0))\log M/\mu \,,
\end{equation}
where $S$ is a finite classical part. More precisely, the geometrical part of the 
above logarithmic divergence, the one coming from the curvature of $S^4$, is supposed to 
be absorbed in the normalization factor $\Z_0$\,, while the part depending on the gauge field 
is indeed a counterterm. 
Using eqs.\ (\ref{phigh}) and (\ref{phigauge}), and rewriting 
$F\cd F$ as in eq.\ (\ref{FF=}), we obtain that 
\begin{equation}\label{bact}
    S_0 = S - \Bigg[ 
    \frac{5}{3}\, \frac{N}{8\pi^2} \int_X\sqrt{g}\,\tr\, F^-\cd\, F^- 
    + \frac{11}{3}\, \frac{N}{8\pi^2} \int_X \tr\, F\wedge F \,
    \Bigg] \log M/\mu \,\,.
\end{equation}
The first counterterm above is effective also in the topologically trivial case and produces 
the renormalization of the field $P$ as discussed in Sect.\ \ref{sect2}, see eq.\ (\ref{redefP}), 
where we find now that $Z_-=-\frac{5}{3}\, \frac{N}{(2\pi)^2} \log M/\mu $, which corresponds to 
\begin{equation}
    \gamma_- = \frac{5}{3}\, \frac{N}{(2\pi)^2} \,\,. 
\end{equation}
This agrees with the result of the computation in Sect.\ \ref{sect2}, see eq.\ (\ref{Gpol}). 

Let us now restrict to the self-dual background, in which case 
we obtain for the measure (\ref{IM1}): 
\begin{equation}\label{IM2}
    dV_k = dv_k\,e^{-S_0(M)}\,M^{\frac{11}{3}Nk}\,R \,,
\end{equation}
where $R$ is $M^2$-independent, but its dimension is such that the whole expression is dimensionless. 
Thus, we can define 
\begin{equation}
    dv_k\,R = d\bar v_k \,\rho^{\frac{11}{3}Nk}\,\bar R \,,
\end{equation}
where $\bar R$ is dimensionless, while $\rho$ is a choice of characteristic size of the instanton field. 
The new `collective coordinate' measure $d\bar v_k$ is dimensionless. For example, in the case when 
$N=2, k=1$, it can be written as $\rho^{-5}d\rho\, d^4x$.

The bare action of eq.\ (\ref{bact}) simplifies to  
\begin{equation}
    S_0 = \tau_0(M)\,k \,,
\end{equation}
with the bare constant $\tau_0(M)$ given by a familiar expression: 
\begin{equation}\label{tau0}
    \tau_0(M) = \ii\theta + \frac{8\pi^2}{g^2} + \frac{11}{3}N\log M/\mu \,,
\end{equation}
where the second term, $8\pi^2/g^2$, is an arbitrary initial real part of $\tau_0$, which we 
included as soon as the latter acquires a renormalization. 
We recognize here the second of two counterterms introduced in eq.\ (\ref{cnt}) with 
$\beta=-\frac{11}{3}N/16\pi^2$.

In terms of renormalized quantities, we can rewrite the instanton measure as 
\begin{equation}\label{IMr}
    dV_k = d\bar v_k \, e^{-\tau k} \bar R \,,
\end{equation}
where now 
\begin{equation}\label{taur}
    \tau(\rho) = \ii\theta + \frac{8\pi^2}{g(\mu)^2} - \frac{11}{3}N\log\rho\mu \,,
\end{equation}
Note that the renormalization generates a $\mu$ dependence 
only in the real part of the constant $\tau$ in front of the topological term in the action (\ref{S}), 
while the imaginary part remains unrenormalized as does a $\theta$-term in the standard \YM\ theory.
The coincidence of this result with a one-loop expression for the instanton measure in the 
standard \YM\ theory (see \cite{NSVZ}) is not accidental. It is explained by the fact that 
the operator $D^*D$, which appears, e.g., in eq.\ (\ref{59}), coincides with the operator 
appearing in the standard YM computation when the background gauge field is self-dual. 
Note, however, that in the \SDYM\ theory there are no higher loop corrections to 
eqs.\ (\ref{IMr}),(\ref{taur}). 
An immediate consequence\footnote{We thank A.\,I.\,Vainshtein for pointing this out.} 
of the latter fact is that the evolution law of the running constant 
$\tau(\mu)=8\pi^2/g(\mu)^2+\ii\theta$ preserves holomorphicity in $\tau$, 
a trivial analogue of the phenomena described in ref.\ \cite{holo}. Indeed, 
given two scales, $\mu_1$ and $\mu_2$\,, we observe that 
$\tau(\mu_1)=\tau(\mu_2)+\frac{11}{3}N\log\mu_1/\mu_2$ is a holomorphic function 
of $\tau(\mu_2)$.

\section*{\large Acknowledgements}
We thank A.\,I.\,Vainshtein, M.\,Bianchi, Ya.\,S.\,Stanev, G.\,Korchemsky, N.\,Nekrasov, and A.\,Gorsky for stimulating discussions. 
The work of A.\,L.\ was accomplished in GNU FNC NIISI RAN program No.6, theme 36.20.
The work of I.\,P.\ (Section 2 of the present paper) was performed at the ITEP with funding from 
Russian Science Foundation (Grant No.16-12-10344).
The work of A.\,R.\ (Section 3 of the present paper) was performed at the IITP with funding from 
the Russian Science Foundation (Grant No.14-60-00150).

\end{document}